\documentclass[journal,letterpaper]{IEEEtran}

\usepackage{graphicx,cite,epsfig,amssymb,amsmath,amsfonts,multicol,subfigure,mathtools,bm,mathrsfs,setspace}
\usepackage{multirow}

\usepackage[numbers,sort&compress]{natbib}

\begin{document}
\title{\huge A Quasi Doppler Method for Signal Transmission to Spatial Perpendicular Directions}
\author{ Bingli JIAO

\thanks{B. Jiao ({\em corresponding author}) and Meng MA are with the Department of Electronics and Peking University-Princeton University Joint Laboratory of Advanced Communications Research, Peking University, Beijing 100871, China (email: jiaobl@pku.edu.cn).}

}

\maketitle

\begin{abstract}

This paper introduces a communication method that can use one information symbol to provide two sets of the independent bits to two receivers in spatial perpendicular directions.  The new communication scheme is realized by switching one signal source among a linear array antenna elements, which are used to emulating a moving transmitter.  The theoretical derivations are presented in the paper.     

\end{abstract}


\IEEEpeerreviewmaketitle

\section{Introduction}
As has been known, Doppler effect is a phenomenon that the frequency would change if there is a relative movement between the transmitter and receiver. The theory has been well established in the early study of electronic magnetic (EM) waves \cite{Jackson1998,Chen2006}.  

To pave the way of developing the proposed method, let us recall Doppler effect in the scenario as shown in Fig.1, which depicts the transmitter moving toward the receiver $R_x$.  The classical Doppler frequency shift can be expressed mathematically 
\begin{eqnarray}
	\begin{array}{l}\label{doppler-01}
	\Delta f = f'-f=\frac{v_x}{\lambda}   \ \ \ for \ \ v_x >0 \ \ or \ \ \ v_x <0
	\end{array}
\end{eqnarray}
where $f'$,$f$, $v_x$ and $\lambda$ are the received frequency at $R_x$, the emitted frequency at $T_{xy}$, the speed of the moving transmitter and the wavelength, respectively. 

\begin{figure}[htb]
	\centering
	\includegraphics[width=0.45\textwidth]{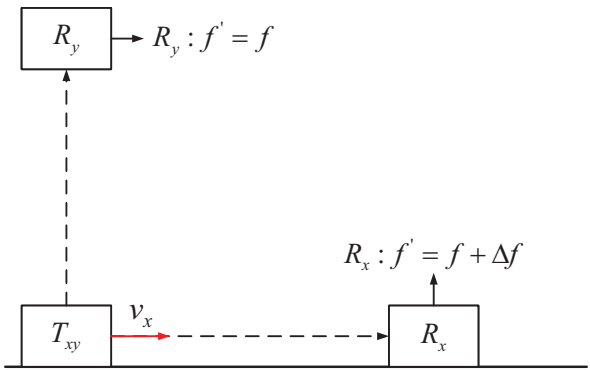}
	\caption{An illustration of Doppler frequency shift.}
	\label{doppler}
\end{figure}

However, if the receiver is located in the perpendicular direction,  there is no Doppler frequency shift, i.e., 
\begin{eqnarray}
\begin{array}{l}\label{doppler-02}
f'=f \ \ \
\end{array}
\end{eqnarray}
ar receiver $R_y$.   

The Doppler effect is essentially found from the electric magnetic EM wave equation 
\begin{eqnarray}
\begin{array}{l}\label{doppler-02}
R_x(t)= Ae^{j\omega t-jK\{x_0-v_x t-x_r+\phi_0\}} \\ 
\end{array}
\end{eqnarray}
with $j=\sqrt{-1}$, where $v_x$ is the moving speed, $R_x(t)$ is the received signal, $A$ is the received amplitude, $\omega$ is the frequency,  $x_0$ is the initial position of the transmitter, $\phi_0$ is the initial phase at the receiver, $x_r$ is the position of the receiver, K is the wave number with $K=2\pi/\lambda$, where  $\lambda$ is the wavelength.  

In wireless communication, the Doppler effect in \eqref{doppler-02} can be expressed as the base band signal  
\begin{eqnarray}
\begin{array}{l}\label{douppler-02}
\hat{R}_x(t)=Ae^{j\Delta \omega t+j\phi_s}  \ \ \ for \ \ \ i-1,2,....
\end{array}
\end{eqnarray}
after the down frequency operation, where $\omega = 2\pi f$, $\Delta \omega = 2\pi \Delta f = K v_x$,  $\hat{R}(t)$ is the base-band signal, and $\Delta \omega t$ can be regarded as the continuous phase shift due to the Doppler effect \cite{Tse2005}.   
   
While, in the perpendicular direction (see Fig.1), there is neither Doppler frequency shift nor phase shift at the receiver $R_y$.       

The abovementioned anisotropy of the wave propagation will be used to the signal modulations to transmit independent information bits to the two receivers with the perpendicular geometry as shown in Fig.1. 

\section{System Model}
To extract the Doppler effect for enabling the proposed method, we use an linear antenna array to replace the moving transmitter for obtaining the Doppler effects. 

To show the key point for the, so called, Doppler phase modulation, we design a simple system as shown Fig.2, in which one can find a harmonic EM signal source and a linear antenna with N elements uniformly distributed along x-axis in range $x_n \in (0, 2\lambda)$, where $\lambda$ is the wave length.  The distance between any two antennas is equal to $d$ and the harmonic signal source in form of $e^{j\omega t}$ is switched among the antenna elements sequentially.  The signal delay between the source and each antenna element is designed same and the residence time at each element is assumed at $T$.  Finally, the switching time is assumed much smaller than that of the residence time.     

We use the light of sight channel model and ignore the multi path signals at the receivers in the following derivations.     

\begin{figure}[htb]
	\centering
	\includegraphics[width=0.45\textwidth]{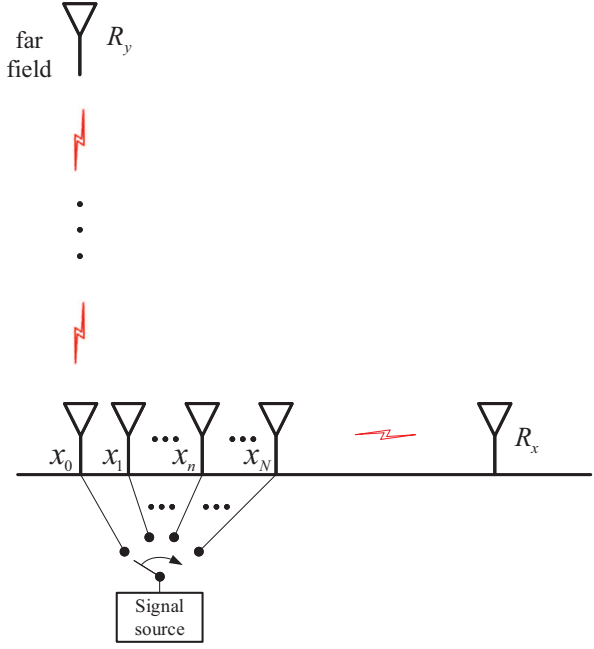}
	\caption{The proposed array system for obtaining the Doppler effects.}
	\label{doppler}
\end{figure}

\subsection{Quasi-Doppler Transmission}
The basic idea to obtain the Doppler effects is to switch the signal source among the antenna 1 to $N$, sequentially, to emulate a discrete movement along the line of the antenna array.   

To analyse the communication in the two perpendicular directions, we define the procedures of the signal source to  $R_x$ as the parallel transmission and that to $R_y$ as the perpendicular transmission.  

Let us work on the parallel transmission, first, by switching the signal source from antenna element 1 to N for emulating a transmitter moving through the positions of each elements at speed of $v_x = d/T$. The received base-band signal at $R_x$ can be written as 
\begin{eqnarray}
\begin{array}{l}\label{douppler-003}
\hat{R}_x(t)= h_xe^{-jKv_x  T \lfloor t/T \rfloor + j\phi_{x}} +n_0 \ \ \
\end{array}
\end{eqnarray}
where  $\lfloor \rfloor$ represents the operation of taking the integer number of low value, $h_x$, $\phi_{x0}$, $n_0$ and $n$ are the amplitude and phase of the channel response, the thermal noise and index of the array elements.  

The above equation shows the Doppler frequency effect by the frequency shift of $Kv_x$ at base-band level with discrete timing. Thus, we refer to this as the Quasi-Doppler effects because the result of \eqref{douppler-003} are obtained by the switching among the antenna elements instead of the moving transmitter.   

One can find that the quasi Doppler effect can be the same as that of conventional Doppler if the value of the distance $d$ approaches to zero, because \eqref{douppler-003} becomes mathematically \eqref{douppler-02} eventually.  

To conclude the above derivations, one can find that the quasi Doppler effect can be essentially same as that of convention one at the base-band signals. While, in the perpendicular direction, $R_y$ receives no any Doppler frequency shift.  

\subsection{Discrete Phase Modulation by Selecting Antenna}
It is interesting to note that \eqref{douppler-003} can be re-written as 
\begin{eqnarray}    
\begin{array}{l}\label{doppler-phase-02}
\hat{R}_x(x_n)=  \sqrt{E_s} e^{ jK x_n + j\phi_0} + n_0\ \ \,  
\end{array}
\end{eqnarray}
because $v_x t_n = x_n$, where $E_s$ is the symbol energy. This indicates that we can realize a phase modulation from  $0$ to $2\pi$ in principle by switching the signal source directly onto one antenna located at the appropriate position along x-axis.  

We define the phase term in \eqref{doppler-phase-02} as the Phase-Doppler (PhD) expressed by 
\begin{eqnarray}    
\begin{array}{l}\label{douppler-phase-03}
\phi_D= K x_n\ \ \,  
\end{array}
\end{eqnarray}
where $x_n$ is the element's position that can be designed by the specified PhD modulation $\phi_D$.  

Since the PhD $\phi_D$ does appear in the perpendicular transmission due to the Doppler effects,  we can separate the PhD phase modulation from that of the perpendicular transmission as explained next.        

\subsection{Joint Signal Modulation Method}
This subsection explains the phase modulations of the parallel- and perpendicular transmission as follows.     

Assume that the transmitter sends a symbol with $e^{j\phi_x}$ to parallel transmission and $e^{j\phi_y}$ to perpendicular transmission, respectively, where the two phases $\phi_x$ and $\phi_y$ are independent of each other.  

Since the PhD modulation does not affect the phase in the perpendicular transmission at $R_y$, we use the conventional signal modulation to at the signal source by $e^{j\phi_y}$ directly.  The received signal can be written as 
\begin{eqnarray}    
\begin{array}{l}\label{douppler-phase-05}
\hat{R}_y = h_y \sqrt{E_s}e^{j\phi_y+\phi_{y0}} + n_0  
\end{array}
\end{eqnarray}
where $\hat{R}_y$ is the received base-band signal, $h_y$ and $\phi_{y0}$ and $n_0$ are the amplitude and phase of the channel response and the thermal nose term respectively.  We note that the perpendicular transmission with \eqref{douppler-phase-05} will not be affected by the PhD modulation due to the property of Doppler effects.   

For the parallel transmission, the phase of the received signal of $R_x$ is the phase summation of the perpendicular transmission and the Dh.D as expressed by    
\begin{eqnarray}    
\begin{array}{l}\label{douppler-phase-06}
\phi_x = \phi_y + \phi_D  
\end{array}
\end{eqnarray}  
when $\phi_x$ is the phase modulation at the parallel transmission.  Thus, $\phi_D$ can be calculated by   
\begin{eqnarray}    
\begin{array}{l}\label{douppler-phase-07}
\phi_D = \phi_y-\phi_x              
\end{array}
\end{eqnarray} 
when for $\phi_y \le \phi_D$, and
\begin{eqnarray}    
\begin{array}{l}\label{douppler-phase-08}
\phi_D = \phi_a + \phi_x    
\end{array}
\end{eqnarray} 
	with $\phi_y + \phi_a =2\pi$, for  $\phi_y >  \phi_D$.  

Then, the parallel transmission can be realized by using PhD.  The base band signal at receiver $R_x$ can be written as 
\begin{eqnarray}    
\begin{array}{l}\label{douppler-phase-011}
\hat{R}_x = h_x \sqrt{E_s}e^{j\phi_x+\phi_{x0}} + n_0  
\end{array}
\end{eqnarray}
where $\phi_x$ is the solution of \eqref{douppler-phase-07} or  \eqref{douppler-phase-08}, $h_x$ and $\phi_{x0}$ are the amplitude and the phase of the channel response of the parallel transmission.  

The plausible feature of the proposed method is found for the one symbol transmission to appear with $\phi_x$ at $\hat{R}_x$ and $\phi_y$ at $\hat{R}_y$, in manner of co-time and co-frequency, simultaneously without any interference between the two.  Thus, both the frequency- and power,  efficiency is doubled in transmission comparison with most conventional methods.

\section{Conclusion}

The proposed method extracts essentials of Doppler effects for the signal transmission in spatial perpendicular directions.  By creating the model of quasi Doppler effect, the discrete modulation method is realized by selecting antenna element to provide the independent phase modulations to the two receivers in spatial perpendicular geometry. The advantages are found for the doubled spectral- and power efficiency with the transmission.

\end{document}